\documentclass[aps,twocolumn,showpacs,superscriptaddress,groupedaddress,nofootinbib,preprintnumbers]{revtex4} 
\pdfoutput=1
\usepackage{graphicx}  
\usepackage{dcolumn}   
\usepackage{bm}        
\usepackage{amssymb}   
\usepackage{amsmath}   
\usepackage{braket}
\usepackage{footnote}
\usepackage{subfigure}
\usepackage{color}
\usepackage{here}
\usepackage{multirow}
\usepackage{enumitem}
\usepackage[utf8]{inputenc}

\usepackage[T1]{fontenc}
\usepackage{hyperref}
\usepackage[svgnames]{xcolor}
\hypersetup{colorlinks=True,
					breaklinks=True,
             		urlcolor=RoyalBlue,
             		citecolor=RoyalBlue,
             		linkcolor=RoyalBlue}

\begin{document}

\title{ New constraints on Heavy Neutral Leptons from Super-Kamiokande data }

\date{\today}

\author{P.~Coloma}
\affiliation{IFIC (CSIC-UVEG), Edificio Institutos Investigaci\'on, 
Apt.\ 22085, E-46071 Valencia, Spain}
\author{P.~Hern\'andez}
\affiliation{IFIC (CSIC-UVEG), Edificio Institutos Investigaci\'on, 
Apt.\ 22085, E-46071 Valencia, Spain}
\author{V.~Mu\~noz}
\affiliation{IFIC (CSIC-UVEG), Edificio Institutos Investigaci\'on, 
Apt.\ 22085, E-46071 Valencia, Spain}
\author{I.~M.~Shoemaker}
\affiliation{Center for Neutrino Physics, Department of Physics, Virginia Tech University, Blacksburg, VA 24601, USA}
\preprint{IFIC/19-51}

\begin{abstract}
Heavy neutral leptons are predicted in many extensions of the Standard Model with massive neutrinos. If kinematically accessible, they can be copiously produced from kaon and pion decays in atmospheric showers, and subsequently decay inside large neutrino detectors. 
We perform a search for these long-lived particles using Super-Kamiokande multi-GeV neutrino data and derive stringent limits on the mixing with electron, muon and tau neutrinos as a function of the long-lived particle mass. We also present the limits on the branching ratio versus lifetime plane, which are helpful in determining the constraints in non-minimal models where the heavy neutral leptons have new interactions with the Standard Model. 
\end{abstract}
\maketitle


\textit{Introduction.} There are compelling reasons to believe that neutrino masses are the first manifestation of a new physics (NP) scale, which can be identified with the mass of the heavy mediator(s) that generates neutrino masses.
Under this assumption, neutrino masses and mixings provide information on a combination of the mediator mass and couplings. Although an upper limit of the new physics scale can be derived by requiring the new couplings to be perturbative, generically, no lower bound results from this constraint. As it is well known, a very high NP scale leads to a hierarchy problem~\cite{Vissani:1997ys,Casas:2004gh}, and to vacuum instability issues~\cite{Casas:1999cd, EliasMiro:2011aa}, both of which can be avoided if the NP scale is not much higher than the electro-weak scale. The possibility that the neutrino mass mediators are light enough to be produced in accelerators and in atmospheric showers is therefore worth exploring.
 
The most popular extension of the Standard Model (SM) realizing neutrino masses is the 
Type-I seesaw model~\cite{Minkowski:1977sc,GellMannRamond,Yanagida:1979as,Mohapatra:1979ia}, with at least two heavy Majorana singlets $N_j$:
\begin{equation}
\label{eq:VNmodel}
 {\mathcal L}_N \supset \sum_j i \bar{N}_j \slash \!\!\!\partial N_j - \left(Y_{\alpha j} \bar{L}_\alpha {\tilde\Phi} N_j + {m_{N_j} \over 2} \bar{N}_j N_j^c \right) + {\rm h.c.} ,
\end{equation}
where $\tilde\Phi \equiv i\sigma_2 \Phi^*$ is the complex conjugate of the Higgs field $\Phi$, $L_\alpha$ is the SM lepton doublet with flavor $\alpha$, $Y$ is a generic Yukawa matrix, and $m_{N_j}$ is the Majorana mass of the singlet $N_j$.   
After spontaneous electro-weak symmetry breaking, the heavy Majorana states mix with the standard neutrinos resulting in a spectrum of three light states with masses $m_\nu \propto m_N^{-1} (Y v)^2$, and two or more heavy neutral leptons (HNL) with masses $\propto m_N$. In this model, all massive neutrino states are admixtures of the standard neutrinos and the singlet states, as dictated by the mixing matrix $U_{\alpha j}$ (which diagonalizes the mass Lagrangian of the whole system). The phenomenology of the HNL depends crucially on their mass and their mixing to the charged leptons. In fact, it is through this mixing that the heavy singlets can be produced either through charged-current (CC) or neutral-current (NC) processes and also how they decay back to SM particles. For simplicity, in this work we will adopt the simplified notation $|U_{\alpha}| \equiv U_{\alpha j}$, assuming that only one of these states is kinematically accessible for our purposes. 

HNLs have been extensively searched for in laboratory experiments, using mainly two types of signatures: either displaced vertices from the decay of the HNL, or through precise determination of the decay product kinematics in meson decays (see, e.g., Refs.~\cite{Atre:2009rg,Bryman:2019ssi,Bryman:2019bjg,Drewes:2015iva} for reviews of available constraints in the MeV-GeV mass range, or Refs.~\cite{SHiP:2018xqw,Ballett:2019bgd,Chun:2019nwi,Dib:2019tuj,Kobach:2014hea,Coloma:2017ppo,Gago:2015vma,Cvetic:2019rms,Cvetic:2018elt,Das:2017rsu,Cvetic:2015naa} for future prospects to improve over current bounds). In Ref.~\cite{Arguelles:2019ziu} we studied in detail the search for long-lived particles produced in the atmosphere, which would then decay inside large-volume neutrino detectors. We derived bounds on HNL for masses above the kaon mass, where laboratory limits are weaker. We found, however, that the limits from atmospheric searches are not competitive with laboratory constraints in this mass range.

In this letter we focus on the lighter mass region instead, where the HNL can be produced in kaon and pion decays for which the atmospheric flux is significantly larger. The most stringent bounds for HNL below the kaon mass come from peak searches \cite{Artamonov:2014urb,CortinaGil:2017mqf} and displaced HNL decay vertex searches~\cite{Bernardi:1985ny,Bernardi:1987ek,Abe:2019kgx}. However, for masses below the kaon and pion mass the HNL becomes very long-lived: although the value of its lifetime in the rest frame ($\tau$) depends strongly on its mass and on its mixing with the light states, in the minimal model described above it ranges between $ c\tau \sim (10^{-4} - 50 ) \times |U_\alpha |^{-2}$~(km), for $ m_N$ between 40~MeV and 400~MeV. This makes atmospheric neutrino detectors a well-suited setup to search for their decay products. 

In this work, we use the framework developed in Ref.~\cite{Arguelles:2019ziu} to extract limits from the multi-GeV muon and electron neutrino data samples observed at Super-Kamiokande (SK). The search of atmospheric HNL in a similar mass range has been considered before for SK~\cite{Kusenko:2004qc,Asaka:2012hc} and IceCube~\cite{Masip:2014xna}. Our analysis significantly improves the methodology of these earlier studies. In addition, extensions of the minimal model of Eq.~(\ref{eq:VNmodel}) that decouple production and decay have been considered in recent works, particularly in relation with the LSND/MiniBoone anomaly~\cite{Bertuzzo:2018itn,Ballett:2018ynz,Arguelles:2018mtc,Coloma:2019qqj,deGouvea:2019qre,Dentler:2019dhz}. In Refs.~\cite{Gninenko:2009ks,Gninenko:2010pr, Magill:2018jla,Fischer:2019fbw} an extension that includes a dipole interaction of the heavy singlets provides a new radiative decay channel, which dominates the HNL decay and significantly reduces its lifetime. Therefore, we will present our limits not only in the context of the minimal HNL model of Eq.~(\ref{eq:VNmodel}) (that is, on the plane $|U_\alpha|$ versus $m_N$), but also on the plane ${\rm Br}(K/\pi\rightarrow N)$ versus $c\tau$, which is phenomenologically motivated. As we will see, this can be useful in order to constrain non-minimal scenarios with uncorrelated production and decay, such as for example the dipole extension of Ref.~\cite{Fischer:2019fbw}.


\textit{HNL production mechanisms.} The leading mechanism for HNL production is through the decays of mesons produced in the atmosphere. The computation is explained in detail in Ref.~\cite{Arguelles:2019ziu}, and here we summarize it for convenience. Defining $\ell$ as the distance traveled between the production point of the HNL to its point of entry in the detector, the HNL production profile (in a differential of distance $d\ell$) can be obtained as:
\begin{eqnarray}
{d \Pi^{\rm decay}_N \over d E d \cos\theta d \ell} = \sum_P \int_{E_P^{\rm min}}^{E_P^{\rm max}} {d E_P\over  \gamma_P \beta_P c \tau_P} {d\Phi_P\over dE_P d\cos\theta }   {d n \over d E  },
\label{eq:master}
\end{eqnarray}
where $dn/dE$ stands for the differential distribution of the HNL energies while $\Phi_P$ is the parent meson flux and, in this work, $P= K^{\pm}, \pi^\pm$.  We have used the Matrix Cascade Equation (MCEq) Monte Carlo software~\cite{Fedynitch:2015zma,Fedynitch:2012fs} to compute the fluxes for the parent mesons in the atmosphere, with the\footnote{While significant variations are expected for the prompt neutrino flux if the cosmic ray or hadronic interaction models are changed (see e.g. Ref~\cite{Fedynitch:2015zma}), the conventional neutrino contribution is understood at the $\mathcal{O}(10-20\%)$ level. We expect similar variations in our results, if these assumptions are modified. } SYBILL-2.3 hadronic interaction model~\cite{Fedynitch:2018cbl}, the Hillas-Gaisser cosmic-ray model~\cite{Gaisser:2011cc} and the NRLMSISE-00 atmospheric model~\cite{Picone:2002go}. In Eq.~\eqref{eq:master}, $\beta_P$
 and $\gamma_P$ are the parent boost factors while $\tau_P$ is its lifetime in the rest frame. For the dominant two-body decay $K^\pm, \pi^\pm \rightarrow N l^\pm_\alpha$ (denoted generally as $P \to N Y$), the differential distribution $dn/dE$ reads
\begin{eqnarray}
{d n ( E_P, E)\over d E  } &=& {1\over \Gamma_P} {d\Gamma(P\rightarrow N Y) \over dE} \nonumber\\
&=&  {{\rm Br}(P\rightarrow N Y) \over \Gamma(P\rightarrow N Y)} {d\Gamma(P\rightarrow N Y) \over dE},
\label{eq:dn2b}
\end{eqnarray}
where Br stands for branching ratio, and 
\begin{eqnarray}
{1 \over \Gamma(P\rightarrow N Y)} {d\Gamma(P\rightarrow N Y) \over dE} = {1\over p_P \sqrt{\lambda(1,y_N^2, y_Y^2)}},
\label{eq:2bd}
\end{eqnarray}
with $y_i \equiv {m_i\over m_P}$ and 
\begin{eqnarray}
\lambda(a,b,c) \equiv a^2+ b^2 + c^2 -2 a b -2 a c -2 b c.
\label{eq:lambda}
\end{eqnarray}
Finally, the kinematical limits for $E_P \gg m_P$  are:
\begin{eqnarray}
E_P^{\rm max(min)} \equiv {2 E\over  (1+ y_A^2-y_Y^2) \mp \sqrt{\lambda(1,y_A^2, y_Y^2)}} .
\end{eqnarray}

\begin{figure}
\begin{center}
\includegraphics[width=0.9\columnwidth]{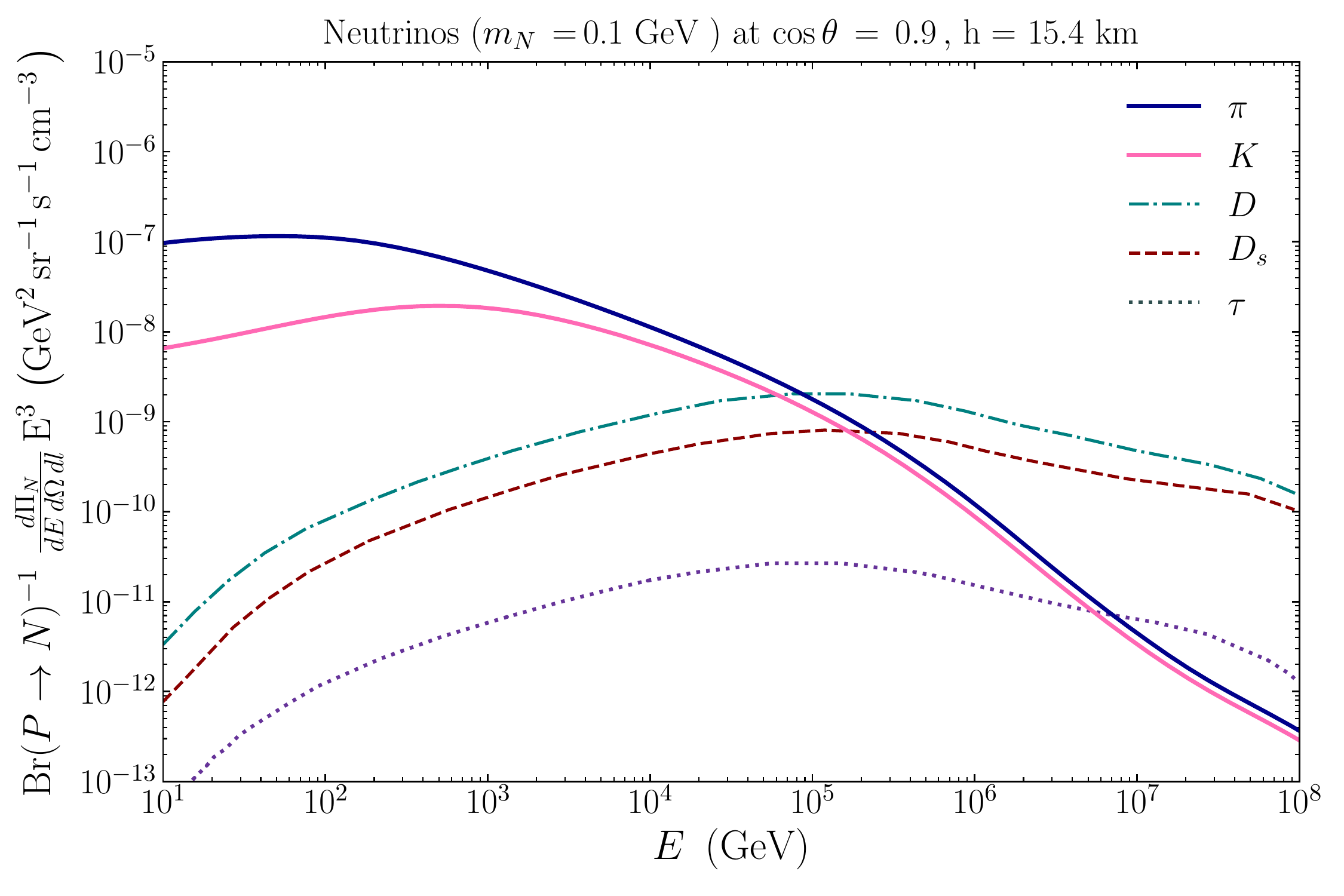} 
\caption{HNL production rate (for $m_N=0.1$ GeV) per energy, solid angle and path length, at a height of 15.4 km and $\cos\theta= 0.9$, for different parent particles $P$. In this work, we consider only $\pi$ and $K$ decays. For comparison we also show the corresponding result for two-body decays of heavier mesons $(D, D_s)$ and leptonic (three-body) $\tau$ decays from Ref.~\cite{Arguelles:2019ziu}. For all lines shown here,  effects due to the mass of the charged lepton produced together with the HNL have been neglected for simplicity. }
\label{fig:kpifluxes}
\end{center}
\end{figure}

Figure~\ref{fig:kpifluxes} shows a representative example of the HNL production from kaon and pion decays in the atmosphere, and compares it to the result obtained for heavier parent mesons and $\tau$ leptons in our previous work~\cite{Arguelles:2019ziu}. As can be seen from this figure, the production profile grows by several orders of magnitude when the mass of the HNL allows for it to be produced from the decays of lighter resonances, due to their more abundant fluxes in the atmosphere.

A second contribution to the flux comes from the HNL production in NC scattering of standard atmospheric neutrinos as they pass through the Earth's matter (for instance in $ \nu \mathcal{N} \to N X$, where $\mathcal{N}$ stands for a nucleon and $X$ is a hadronic shower). This contribution can be estimated as follows. Assuming a flux of standard neutrinos $\phi_\nu$, the HNL produced at a distance $r$ from SK is:
\begin{eqnarray}
{d \Pi^{\rm int}_N \over d \cos\theta d E d r} = \int_{E_\nu^{\rm min}}^{E_\nu^{\rm max}} \!\!\!dE_\nu {d \phi_\nu (r)\over d \cos\theta d E_\nu} {d \sigma_{\nu\rightarrow N}(E_\nu) \over d E} N_A\rho_{\oplus},\nonumber\\
\end{eqnarray}
where $\sigma_{\nu\rightarrow N}$ is the HNL production cross section (which can be estimated as the standard $\nu$ NC scattering cross section  multiplied by the corresponding mixing $|U_\alpha|^2$), while $N_A$ is the Avogadro number and $\rho_{\oplus}$ is the Earth density at distance $r$. This assumes that both standard neutrinos and the resulting HNL are highly boosted and therefore the zenith angle of the HNL is roughly the same as that of the incoming neutrino. The upper and lower limits in the integral correspond to the kinematically allowed ranges, where a SM neutrino scattering on a nucleon at rest can produce a HNL with energy $E$.


\textit{HNL decays in Super-Kamiokande.} The flux that arrives to the detector for HNL produced in meson decays, $\Phi^{\rm decay}_N$, is obtained integrating over all LLPs produced at different distances $\ell$, weighted by their corresponding survival probabilities, as 
\begin{eqnarray}
{d \Phi^{\rm decay}_N\over d\cos\theta dE} = \int_0^{\ell_{\rm max}} d \ell {d \Pi^{\rm decay}_{N}\over d\cos\theta dE d\ell} 
~e^{-{\ell\over L_{\rm decay}}},
\label{eq:fluxA}
\end{eqnarray}
where $L_{\rm decay}$ is the decay length of the HNL in the laboratory frame. The maximum distance $ \ell_{\rm max} \equiv \ell(h_{\rm max},\theta)$ is a simple function of  the maximum height of the atmosphere where cosmic showers are produced, $h_{\rm max} \simeq 80$~km, and the zenith angle. Analogously, the flux entering the detector for HNL produced in the interaction of SM neutrinos in the Earth, $\Phi^{\rm int}_N$ is:
\begin{eqnarray}
{d \Phi^{\rm int}_N \over d\cos\theta d E} = \int_0^{R_{\rm max}(\theta)} dr  {d \Pi_N \over d\cos\theta d E d r} e^{-r/L_{\rm decay}},
\end{eqnarray}
where $R_{\rm max}(\theta) = 2 R_{\oplus} \cos\theta$ is the maximum distance traveled through the Earth for trajectories with a zenith angle $\theta$, $R_\oplus$ being the Earth's radius.

The total number of HNL decays inside the detector, within a given time window $\Delta T$, energies in the interval $[E, E + dE]$ and zenith $[\cos\theta, \cos\theta + d\cos\theta]$, can be computed as
\begin{equation}
\frac{d N}{dE d\cos\theta} = \Delta T  A^{\rm eff}_{\rm decay}(E,\cos\theta) {d \Phi_N\over dE d\cos\theta },
\end{equation}
where $ \Phi_N \equiv \Phi^{\rm int }_N + \Phi^{\rm dec}_N$, and $ A^{\rm eff}$ is an effective area which accounts for the probability that a decay takes place inside the detector. This area can be estimated integrating the surface of the detector normal to the flux direction, weighted by the decay probability of the $N$ inside the detector:
\begin{eqnarray}
A^{\rm eff}_{\rm decay}(E,\cos\theta) = \int dS_{\perp} \left[ 1- e^{-  \Delta\ell_{det} (\cos\theta)\over L_{\rm decay}(E) }  \right] \, .
\end{eqnarray}
Here $\Delta\ell_{det}$ is the length of the segment of the HNL trajectory that cuts into the detector (for explicit expressions we refer the reader to Ref.~\cite{Arguelles:2019ziu}). A cylindrical geometry for SK with height of 40~m and radius of 20~m is assumed.

The two contributions to the total number of events (coming from meson decays and from SM neutrino interactions in the Earth) have a very different angular dependence: while the flux from decays is expected to be larger from above, those from interactions come obviously from below. However, we have checked that the final contribution to the number of events coming from HNL produced in meson decays is several orders of magnitude larger than the one obtained from interactions of SM neutrinos on the Earth. The ratio between the two contributions decreases for larger $c\tau$ but, within the range $c\tau\in [1, 10^4]$~km, it lies within the range $[10^6, 10^2]$. We can therefore safely neglect the contribution from neutrino interactions in the Earth and, in the rest of this work, we will only consider HNL production from meson decays.

In order to derive our limits, we use the data samples as well as the expected neutrino background prediction from Fig. 5 of ref.~\cite{Abe:2017aap}. We use both the $\mu$- and $e$-like fully-contained multi-GeV events, adding the single- and multi-ring samples together (labeled as ``multi-GeV'' and ``multi-ring'' in Fig.~5 in Ref.~\cite{Abe:2017aap}, respectively). In the case of $e$-like events, we add both the $\nu_e$ and $\bar\nu_e$ samples as well. In Ref.~\cite{Abe:2017aap} data are binned in zenith angle, while information on the energy of the events is not publicly available. Therefore, in this work the data is binned only in $\cos\theta$. As for the neutrino energies considered, we integrate over all energies between 1~GeV and 90~GeV, as this is the range corresponding to the fully-contained sample (see Fig.~6 in  Ref.~\cite{Abe:2017aap}). We think this is conservative, as SK may be sensitive to events outside this range. 

The number of events observed in a given sample will also depend on detector efficiencies and reconstruction effects, which should be included in the form of migration matrices giving the relationship between true and reconstructed variables. Such information is however not publicly available for SK. Thus, here we make the simplifying assumption that the efficiencies are independent on the neutrino energy and, in particular, we take a flat detection efficiency $\epsilon^{\alpha} = 0.75$ both for $\mu$ and $e$-like, in line with the values quoted in Ref.~\cite{Abe:2017aap} for the multi-GeV $\nu_e$ event sample. While a priori some loss of efficiency could be expected at high energies (mainly due to a reduction in the containment of the events), we think that the impact on our results would be small since we expect our sensitivities to come mostly from the events at low energies. This is because the heavy neutrino flux produced in the amtosphere will follow a very steep power law, peaked at low energies as shown in Fig.~\ref{fig:kpifluxes}, where we expect the SK efficiencies to be best. We also assume that the angular reconstruction is much better than the width of the bins in zenith angle, so migration between different bins in $\cos\theta$ can be neglected. If this assumption were to be relaxed, our sensitivities would probably be worsened. This is so because in the case of very long-lived particles (as is the case for HNL with masses below 500~MeV, and weakly coupled to the SM) their very long lifetimes lead to a higher sensitivity for specific angular bins, where the distance traveled by the HNL is comparable to (or smaller than) its lifetime in the laboratory frame. While here we have used only publicly available information, a detailed evaluation by the experimental collaboration is needed to validate our results. Finally, we should also point out that we expect an increase in sensitivity if the analysis were to be carried out using both energy and angular information. This is beyond the scope of this work.  

The number of events in the $i$-th bin in $\cos\theta$ in a given sample can therefore be computed as:
\begin{eqnarray}
 &&N^{\rm SK, \alpha-like}_i ={\rm Br}(N \rightarrow \alpha{\rm-like}) \times \nonumber\\
  \!  \! 
&&\int_{1~\rm{GeV}}^{90~\rm{GeV}} dE 
\epsilon^{\alpha}(E) \int_{\cos\theta_{i}^{\rm min}}^{\cos\theta_i^{\rm max}} \! d\cos\theta \frac{dN}{dE d\cos\theta} \, , \nonumber \\
\label{eq:nevents}
\end{eqnarray}
where $\cos\theta_{i}^{\rm min}$ and $\cos\theta_{i}^{\rm max}$ are the lower and upper limits of the bin. Here, ${\rm Br}(\alpha \rm{-like})$ stands for the total branching ratio for all decay channels including muons, electrons or photons in the final state, depending on the sample ($\alpha$) considered. For example, in the case of $\mu$-like we consider only those decay channels including one or more muons in the final state. In the case of $e$-like events we require that no muons are present, but we also include decay channels with photons as these are easily mis-identified with electrons at SK (such as $N \to \nu \pi^0$, since the $\pi^0$ decays promptly to two photons). 


\textit{Results.} In this section we derive limits on HNL production by performing a $\chi^2$ fit to the SK data. A Poissonian $\chi^2$ function has been used:
\begin{eqnarray}
\chi^2 =2  \sum_{i, \alpha}  N_i^{\alpha}+ B_i^{\alpha} - n_i^{\alpha}+n_i^{\alpha} \log\left({n_i^{\alpha}\over N_i^{\alpha}+ B_i^{\alpha}}\right),
\end{eqnarray}
where the sum runs over the angular bins, and $\alpha=\{e$-like, $\mu$-like$\}$. Here, $n_i^\alpha$ stands for the data observed in each bin while $N_i^\alpha$ is the predicted number of signal events and $B_i^\alpha$ is the background prediction, which includes the predicted number of atmospheric neutrino events in the SM. In our $\chi^2$ calculations, we consider separately the $e$- and $\mu$-like samples and we add their two contributions to the total $\chi^2$. Therefore, our limits will be derived using 20 degrees of freedom, corresponding to the total number of bins in $\cos\theta$. In our calculations we find, however, that the sensitivity is largely dominated by the $e$-like contribution since the size of the branching ratio ${\rm Br}(e-\rm{like})$ is much larger than $\rm Br( \mu-like)$ in the mass range considered.

\begin{figure}[htb!]
\begin{center}
\includegraphics[width=1.1\columnwidth]{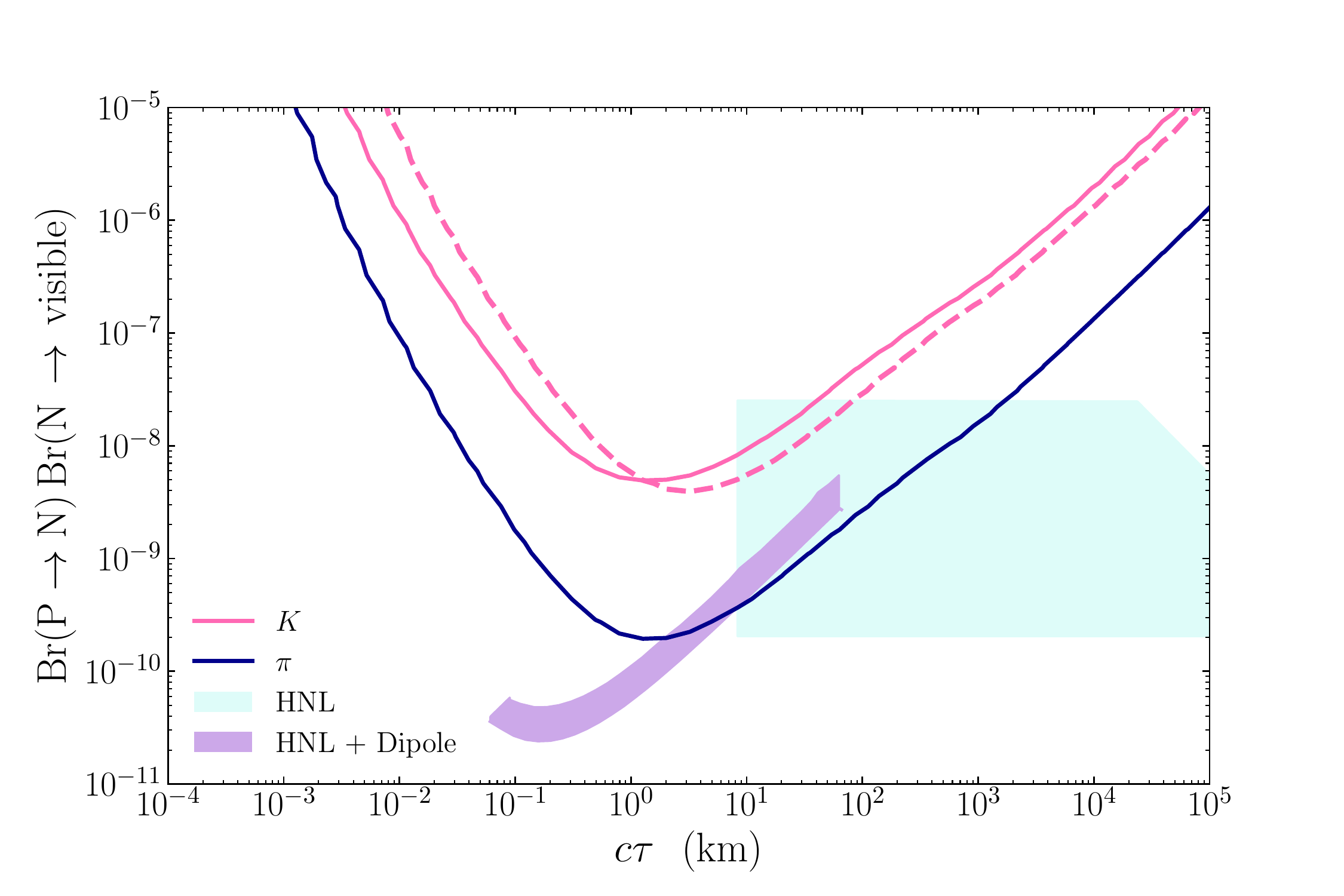} 
\end{center}
\caption{Limits at 90\% confidence level (C.L.), on the plane ${\rm Br}(K/\pi\rightarrow N) \times {\rm Br}(N \to {\rm visible})$ versus $c\tau$ for $m_N= 0.1$ GeV (solid) and $m_N=0.25$ GeV (dashed), for HNL produced in $\pi$ (dark blue) or $K$ (pink) decays. Our limits have been obtained assuming uncorrelated production and decay mechanisms for the HNL. The shaded light blue region corresponds to the (correlated) values obtained for the minimal model, varying the mixings within their presently allowed constraints as explained in the text. The purple area corresponds to the $1\sigma$ allowed region where a HNL with a dipole interaction could explain the MiniBooNE anomaly, according to Ref.~\cite{Fischer:2019fbw}, for $m_N=260$~MeV.}
\label{fig:uncorr}
\end{figure}

First we show in Fig.~\ref{fig:uncorr} the results on the plane ${\rm Br}(K/\pi \rightarrow N)\times {\rm Br}(N\rightarrow {\rm visible})$, where ${\rm Br}(N \to {\rm visible})$ accounts for the probability that the HNL decays visibly in the detector. Our results are presented as a function of $c\tau$, assuming no correlation between the production and decay mechanisms for the HNL. It is interesting to note that SK can outperform the powerful displaced-decay search limits from beam dumps such as those in Ref.~\cite{Bernardi:1985ny,Bernardi:1987ek,Abe:2019kgx}, for models where the decay does not involve two charged tracks, since all laboratory searches request this condition to reduce background contamination. In the extended model of Ref.~\cite{Fischer:2019fbw} this is precisely the case, since the HNL decay is dominated by the radiative $N\rightarrow \nu \gamma$ decay via the dipole interaction, therefore the stringent limits from PS191 \cite{Bernardi:1985ny,Bernardi:1987ek} and the recent T2K limits~\cite{Abe:2019kgx}  do not apply.  The shaded purple region in Fig.~\ref{fig:uncorr} shows the range where the MiniBooNE anomaly could be explained, for a HNL with $m_N=260$ MeV, extracted from Ref.~\cite{Fischer:2019fbw}. In this case, the HNL would only be produced in kaon decays and, unfortunately, our limits from $K^\pm$ decay (pink lines) fall short to probe this region. However, SK would be sensitive to non-minimal models with HNL produced in $\pi^\pm$ decays with a similar value of the production BR and lifetime and, obviously, larger neutrino experiments such as DeepCore or Hyper-Kamiokande could significantly improve over these constraints. 

In addition, the shaded light blue region in Fig.~\ref{fig:uncorr} shows the expectation for the minimal model outlined in Eq.~(\ref{eq:VNmodel}), for $m_N=250$ MeV and mixing matrix elements in the range $|U_{e/\mu}|^2 \in [10^{-8},10^{-10}]$, $|U_\tau|^2 \in [10^{-10}, 10^{-4}]$ (in agreement with current constraints from Ref.~\cite{Atre:2009rg}). As readily seen from this figure, relevant constraints are expected in this case. In view of this result, next we derive constraints on the minimal scenario, assuming only one non-vanishing $U_\alpha$ at a time. 
\begin{figure*}[t!]
\begin{center}
\includegraphics[width=1.\columnwidth]{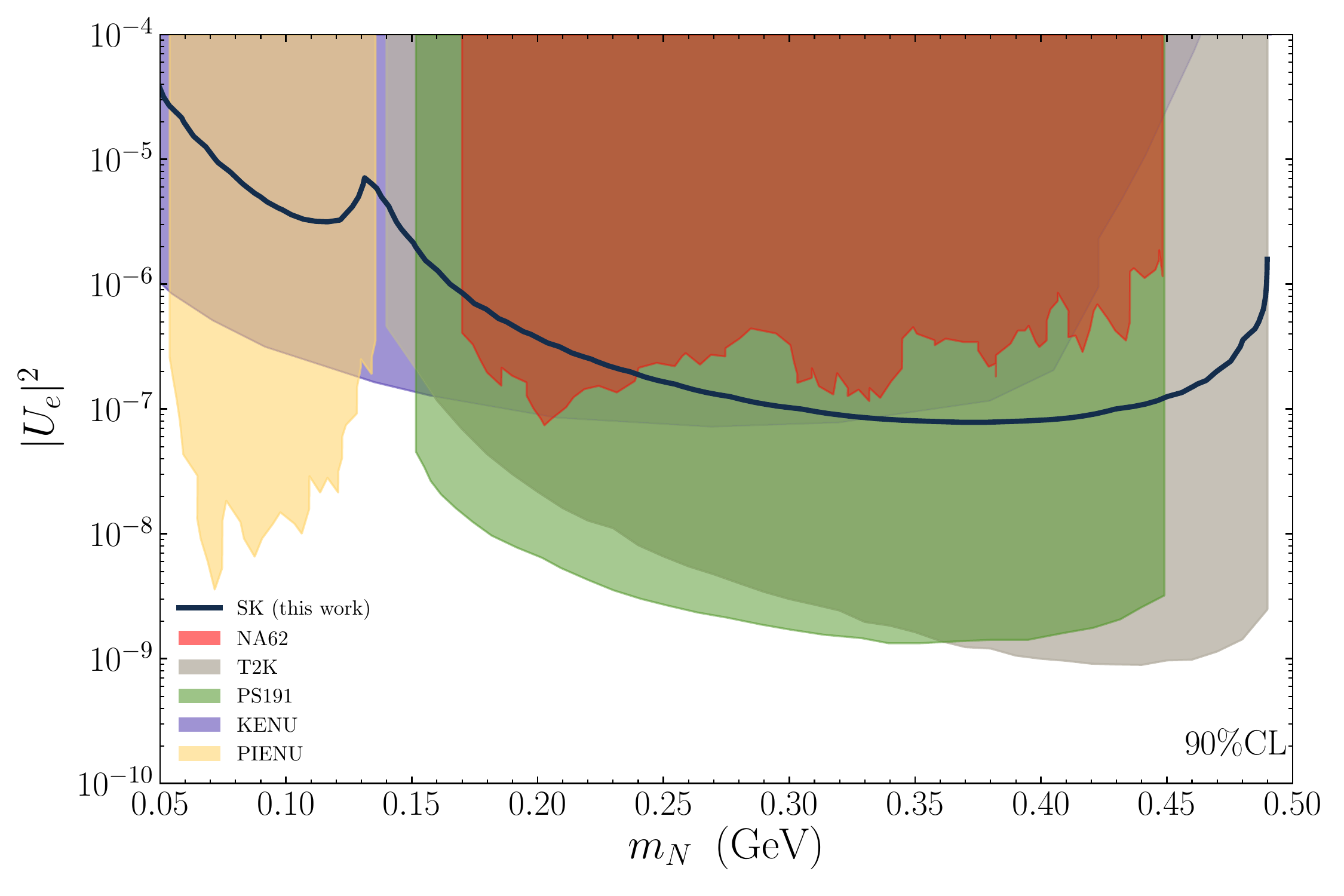} 
\includegraphics[width=1.\columnwidth]{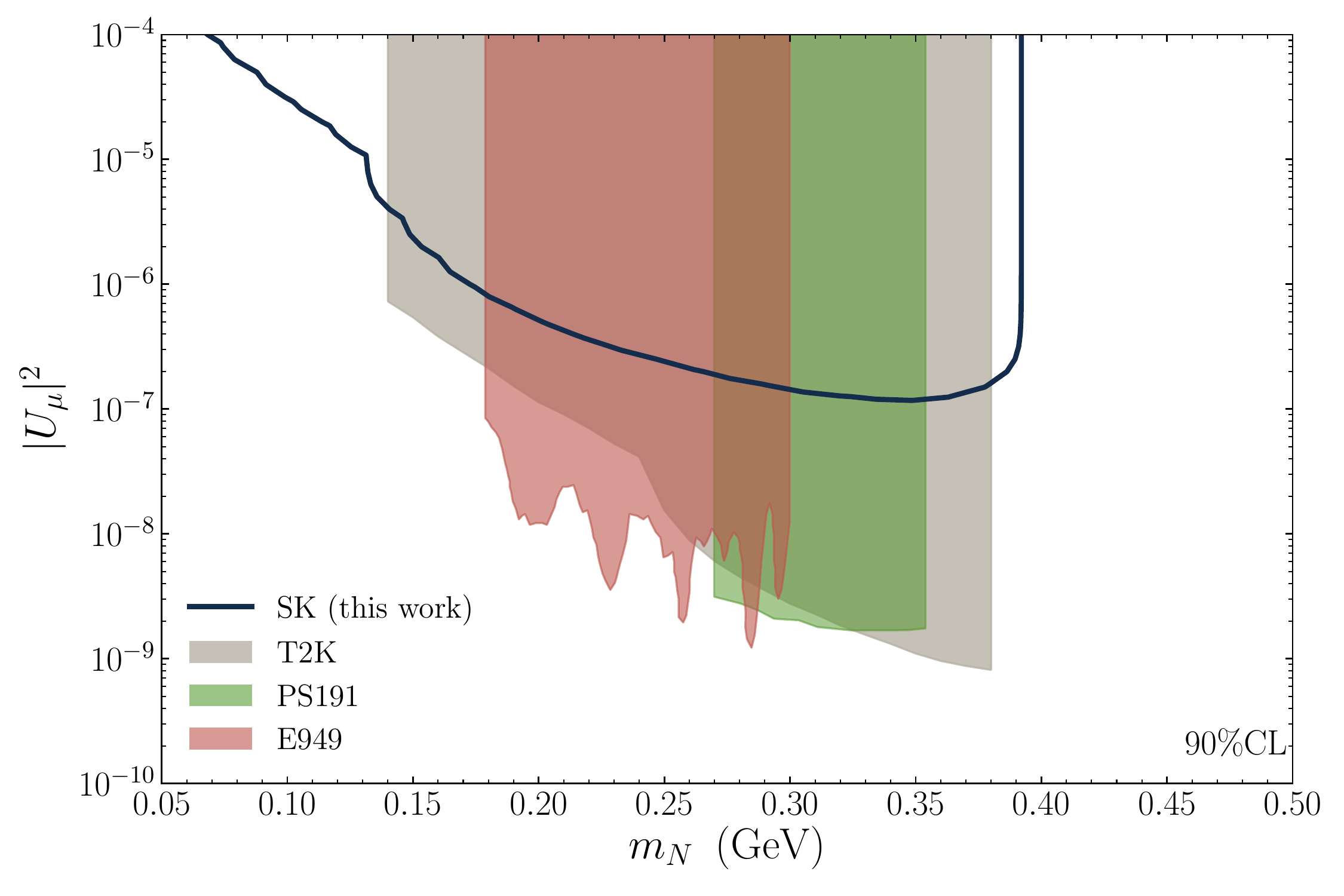} 
\end{center}
\caption{ SK constraints on the minimal HNL scenario at 90\% C.L., projected onto the plane $|U_e|^2$ vs $m_N$ (for $U_{\mu,\tau}=0$). Our results (solid black lines) are compared to corresponding limits obtained for the T2K near detector~\cite{Abe:2019kgx}, NA62~\cite{CortinaGil:2017mqf}, E949~\cite{Artamonov:2014urb}, PS191~\cite{Bernardi:1985ny,Bernardi:1987ek}, and PIENU~\cite{Mischke:2018qmv}. The line labeled as KENU was derived in Ref.~\cite{Bryman:2019bjg} using precision measurements of leptonic decay channels of the kaon~\cite{Lazzeroni:2012cx}. }
\label{fig:Uelimit}
\end{figure*}
Our results are shown in Fig.~\ref{fig:Uelimit} for $|U_e|^2$ (left panel) and for $|U_\mu|^2$ (right panel), at 90$\%$ confidence level (C.L.). In the case of $|U_e|^2$ the contribution from $\pi^\pm$ decays is clearly dominant for $m_N < 140$~MeV, as can be seen from the peak in sensitivity at around 0.1~GeV. We find that the limits derived from our simplified analysis is already able to set tight constraints on the mixing of HNL with the electron and muon neutrino sectors, between $10^{-6}$ and $10^{-7}$ for $m_N$ in the range between 150~MeV and 450~MeV. Our limits are also compared with those obtained from displaced decay searches at PS191~\cite{Bernardi:1985ny,Bernardi:1987ek} and at the T2K near detector~\cite{Abe:2019kgx}, as well as from peak searches in E949~\cite{Artamonov:2014urb}, PIENU~\cite{Mischke:2018qmv}, and NA62~\cite{CortinaGil:2017mqf}. We also show the resulting bound derived in Ref.~\cite{Bryman:2019bjg} from the measurement of the kaon decays into electrons or muons~\cite{Lazzeroni:2012cx}. In the case of $|U_e|^2$, the limits obtained from SK are comparable or even better than analogous limits from peak searches, while they are not competitive with those from displaced decay searches. In the case of $|U_\mu|^2$, peak searches in E949 also yield better constraints than our limits from SK data. 

\begin{figure}
\begin{center}
\includegraphics[width=1.\columnwidth]{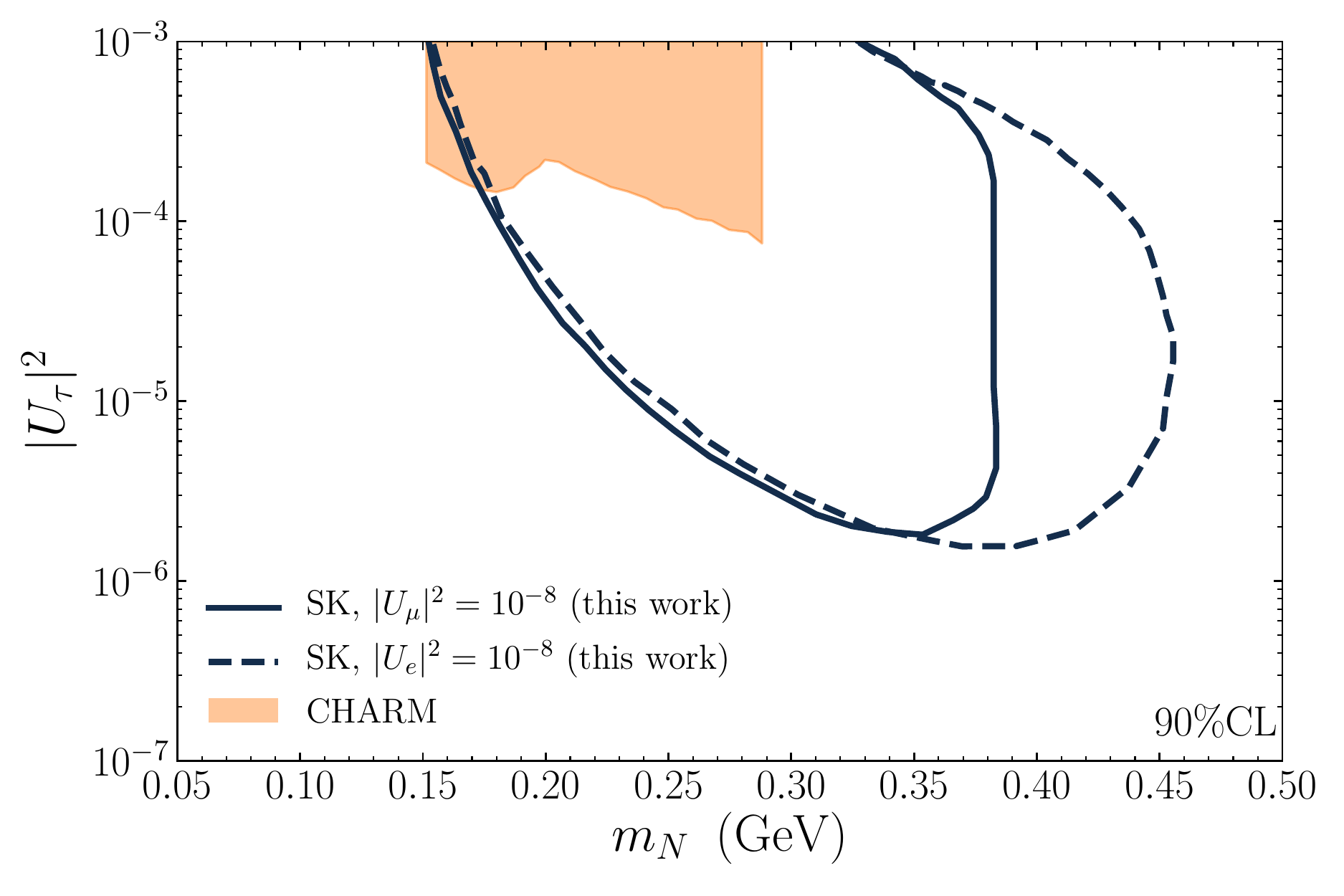} 
\end{center}
\caption{ SK constraints on the minimal HNL scenario at 90\% C.L., projected onto the plane $|U_\tau|^2$ vs $m_N$, for $|U_{e}|^2=10^{-8}$ and $|U_\mu|^2=10^{-8}$. Our results (black lines) are compared to the limits from CHARM~\cite{Orloff:2002de} (which however have been obtained under the assumption $|U_e|^2 = |U_{\mu}|^2 = 0$). }
\label{fig:Utaulimit}
\end{figure}

Finally, while the HNL cannot be produced via $|U_\tau|$ in $K$ or $\pi$ decays (because it is not possible to produce the HNL together with a $\tau$ lepton in this case), competitive limits can still be derived on $U_\tau$ if we allow for non-vanishing $|U_\alpha|,~ \alpha=e,\mu$, even if these are well below present bounds from laboratory experiments. The reason is that, in this case, the HNL could be produced via the mixing $U_e$ or $U_\mu$, and a large $U_\tau$ can induce a significant decrease in its lifetime of the HNL while allowing for a significant branching ratio into $e$-like or $\mu$-like events through NC-mediated decays. Therefore, in Fig.~\ref{fig:Utaulimit} we show the sensitivity to $|U_\tau|$, as a function of $m_N$, for fixed $U_e=10^{-8}$ or $|U_\mu|=10^{-8}$ (which are both below the best present upper bounds). Our limits obtained in this way are already much better than existing direct constraints on $|U_\tau|$ from CHARM~\cite{Orloff:2002de}, which however have been obtained assuming vanishing values for $|U_e|$ and $|U_\mu|$. A similar exercise can be done for PS191, which would probably lead to better limits on $|U_\tau|$ than the SK results, for the same assumed values of $|U_e|$ and $|U_\mu|$. 

\textit{Conclusions.} In this letter, we have used the latest publicly available SK data to derive strong constraints on HNL production from kaon and pion decays in the atmosphere. Using a $\chi^2$ analysis, and binning our events only in $\cos\theta$, we find that SK data is able to provide strong constraints on the minimal HNL scenario for masses between 150~MeV and 400~MeV. It is therefore expected that a more detailed analysis performed by the collaboration may be able to significantly improve over our results. We have also shown our limits in the Br vs $c\tau$ plane, which is applicable to a wider range of NP models where the HNL interact with the SM not only via mixing but also through other interactions (such as, for instance, a dipole moment). Finally, we have used our results to show how, in the case of non-vanishing $U_e$ or $U_\mu$ well below current constraints from laboratory searches, SK data could be used to set strong bounds on $U_\tau$, well below the direct limits presently available from CHARM data. We expect a similar improvement of PS191 bounds under similar assumptions, though.
\\
\\

\begin{acknowledgments}

PC thanks the CERN Theory Division for support and hospitality, and the Fermilab Theory Group for their hospitality during completion of this work. The work of VM is funded by CONICYT PFCHA/DOCTORADO BECAS CHILE/2018 - 72180000. This work was partially supported by grants FPA2017-85985-P, PROMETEO/2019/083, and  the European projects H2020-MSCA-ITN-2015//674896-ELUSIVES and 690575-InvisiblesPlus-H2020-MSCA-RISE-2015. The work of I.M.S. is supported by the U.S. Department of Energy under the award number \protect{DE-SC0020250}. 
\end{acknowledgments}

\bibliographystyle{apsrev-1}
\bibliography{biblio.bib}

\end{document}